\documentclass[journal]{IEEEtran}

\usepackage{amsmath, amsfonts, amssymb}
\usepackage{algorithmic}
\usepackage{algorithm}
\usepackage{array}
\usepackage[caption=false,font=normalsize,labelfont=sf,textfont=sf]{subfig}
\usepackage{textcomp}
\usepackage{stfloats}
\usepackage{url}
\usepackage{verbatim}
\usepackage{graphicx}
\usepackage{cite}

\begin{document}

\title{Symbol-Level GRAND for High-Order Modulation over Flat Fading Channels}

\author{Ioannis Chatzigeorgiou,~\IEEEmembership{Senior Member,~IEEE,} and Francisco A. Monteiro,~\IEEEmembership{Member,~IEEE}
        % <-this % stops a space
\thanks{I. Chatzigeorgiou is with the School of \mbox{Computing \& Communications}, Lancaster University, UK, e-mail: i.chatzigeorgiou@lancaster.ac.uk.}% <-this % stops a space
\thanks{F.~A.~Monteiro~is~with~Instituto~de~Telecommunica\c{c}\~{o}es,~and~\mbox{ISCTE-Instituto} Universit\'{a}rio de Lisboa, Portugal, e-mail: francisco.monteiro@lx.it.pt.}}


% % column for its text to clear the IEEEpubid mark.

\newtheorem{proposition}{Proposition}

\maketitle

\begin{abstract}
Guessing random additive noise decoding (GRAND) is a noise-centric decoding method, which is suitable for ultra-reliable low-latency communications, as it supports high-rate error correction codes that generate short-length codewords. GRAND estimates transmitted codewords by guessing the error patterns that altered them during transmission. The guessing process requires the generation and testing of error patterns that are arranged in increasing order of Hamming weight. This approach is fitting for binary transmission over additive white Gaussian noise channels. This letter considers transmission of coded and modulated data over flat fading channels and proposes a variant of GRAND, which leverages information on the modulation scheme and the fading channel. In the core of the proposed variant, referred to as symbol-level GRAND, is an analytical expression that computes the probability of occurrence of an error pattern and determines the order with which error patterns are tested. Simulation results demonstrate that symbol-level GRAND produces estimates of the transmitted codewords notably faster than the original GRAND at the cost of a small increase in memory requirements.
\end{abstract}


% ---------- Introduction ----------

\section{Introduction}
\label{sec:intro}

\IEEEPARstart{T}{he} requirement for ultra-reliable low latency communication (URLLC) was introduced in fifth generation (5G) networks in order to support services that have stringent requirements for extremely low latency (e.g., $1$ ms) and high reliability (e.g., $99.999\%$). Examples of applications that rely on URLLC include machine-type communications for the industrial internet of things (IIoT), virtual reality and driverless vehicles~\cite{Nouri2020, Monteiro2021}. Ultra reliability and low latency imply the use of high-rate error correction codes that generate short codewords. However, in an effort to achieve Shannon's capacity, emphasis in pre-5G systems was placed on the construction of codes that map information words onto long codewords. Codes that were invented in the 1960s, e.g., BCH codes and Reed-Solomon codes, have sparked renewed interest~\cite{Pfeifer2017} for URLLC use cases; this is because they can assign short codewords to information words, but they support only a limited number of code rates. On the other hand, random linear codes (RLCs) can support any code rate but RLC decoding is an NP-hard problem~\cite{Becker2012}, and is therefore considered impractical.

The recently proposed guessing random additive noise decoding (GRAND) \cite{Duffy2019} has enabled universal decoding, that is, it can decode any linear code, including RLCs. GRAND leverages the fact that noise entropy decreases as the channel conditions improve; therefore, the list of all possible error patterns that could alter a transmitted codeword reduces in size. Attempting to guess the most likely error pattern becomes more efficient than searching for the most likely transmitted codeword in the code-book, and yields maximum-likelihood performance. A code-book membership test is required to verify whether an error pattern corresponds to a valid codeword. GRAND arranges prospective error patterns in descending order of likelihood, runs the membership test on each error pattern, and returns the first error pattern that passes the test.

Error correction codes are usually combined with \mbox{high-order} modulation schemes in wireless systems in order to improve spectral efficiency. At the receiver, a demodulator converts the sequence of received modulated symbols into a stream of bits. The original GRAND attempts to estimate the transmitted codeword from the input stream of bits but knowledge of the modulation type is not exploited in the search for error patterns that satisfy the conditions for code-book membership. The objective of this letter is to develop a variant of GRAND that treats the input stream of bits as an equivalent sequence of hard-detected modulated symbols corrupted by flat fading and additive noise. We refer to the proposed decoder as \textit{symbol-level} GRAND. The modulation scheme considered in this work is $M$-ary quadrature phase modulation ($M$-QAM).

A modification of GRAND that leverages information on the adopted modulation scheme was recently proposed by Wei An~\textit{et al.}  \cite{An2022}. Their paper considers the position of $M$-QAM symbols on the constellation plane, obtains the probability of a symbol transitioning to one of its nearest neighbors, introduces this probability into a Markov chain that models a channel with memory, and identifies rules that place constraints on the generation of error patterns.
In contrast to~\cite{An2022}, we derive a closed-form expression for the probability that the input stream of bits is a sequence of a particular combination of binary strings that correspond to symbols in different positions of the constellation diagram and are, thus, in the proximity of different numbers of nearest and next-nearest neighbors. This probability expression is then used as the basis for the generation of error patterns in order of likelihood, when transmission is over a flat fading channel. In essence, both our letter and~\cite{An2022} design algorithms that consider the modulation scheme in the decoding process but the estimation and ordering of the error patterns follow different rules because they have been tailored to different channel models. Abbas \textit{et al.}  \cite{Abbas2022} also look into GRAND for flat fading channels but do not consider high-order modulation; the focus of their work is on the design of a variant of GRAND that exploits receive diversity.

In the remainder of this letter, Section~\ref{sec:system_model} describes the system model, Section~\ref{sec:SL_GRAND} formulates the problem and develops symbol-level GRAND, Section~\ref{sec:results} presents and discusses results, and Section~\ref{sec:conclusion} summarizes key conclusions.

% ---------- System Model ----------

\section{System Model}
\label{sec:system_model}

Consider a coding and modulation scheme that employs an $[n,k]$ binary linear block code followed by square \mbox{Gray-coded} \mbox{$M$-QAM}. The $[n,k]$ block code uses a $k\times n$ generator matrix $\mathbf{G}$ to encode a $k$-bit input information word, represented by $\mathbf{u}\in\{0,1\}^k$, into an $n$-bit codeword $\mathbf{x}=\mathbf{u}\mathbf{G}$ for $n>k$. The set of all $2^k$ valid codewords is known as the code-book $\mathcal{C}\subseteq\{0,1\}^n$ of the $[n,k]$ code. The $n$ bits of codeword $\mathbf{x}$ are divided into $L$ strings of $\log_2\!M$ bits, i.e., $L=n/\log_2\!M$. If $E_b$ denotes the energy per information bit, then $(k/n)E_b$ is the energy per codeword bit, and $(\log_2\!M)(k/n)E_b$ is the energy per string of $\log_2\!M$ bits. \mbox{$M$-QAM} maps the $L$ strings onto $L$ symbols of a complex set $\mathcal{S}\subset\mathbb{C}$, which has cardinality $M$ and average energy per symbol $(\log_2\!M)(k/n)E_b$. The sequence of $L$ symbols, denoted by $\mathbf{s}\in\mathcal{S}^L$, is transmitted to a receiver over a single-input single-output (SISO) channel.

The channel is impaired by additive white Gaussian noise (AWGN) and flat fading. The relationship between the input and output of the discrete-time equivalent channel is
\begin{equation}
\label{eq:channel_in_out}
\mathbf{r}=h\mathbf{s}+\mathbf{z},
\end{equation}
where $\mathbf{r}\in\mathbb{C}^L$ is the sequence of noisy $M$-QAM symbols at the output of the channel, and $\mathbf{z}\in\mathbb{C}^L$ is a sequence of $L$ zero-mean, mutually independent, complex Gaussian random variables with variance $N_0$. The fading coefficient $h$ is a zero-mean complex random variable with variance $\mathrm{E}\bigl[\lvert h^2 \rvert\bigr]=1$. The value of $h$ remains constant during the transmission of sequence $\mathbf{s}$ but changes independently from one sequence to the next. If $\gamma\!\triangleq\!(\log_2\!M)(k/n)(E_b/N_0)$, the \textit{instantaneous} signal-to-noise ratio (SNR) at the receiver is given by $\lvert h \rvert^2 \gamma$, while the \textit{average} SNR takes the form $\mathrm{E}\bigl[\lvert h \rvert^2 \gamma\bigr]=\mathrm{E}\bigl[\lvert h \rvert^2 \bigr]\gamma=\gamma$.

We assume that channel state information is available at the receiver, i.e., the value of $h$ is known for every received sequence, and coherent detection is possible. The received sequence $\mathbf{r}$ is multiplied by $h^{-1}\triangleq h^{*}/\lvert h\rvert^2$, where $h^{*}$~and~$\lvert h\rvert$ denote the complex conjugate and magnitude, respectively, of $h$. The product $h^{-1}\,\mathbf{r}$ is input to a hard-detection \mbox{$M$-QAM} demodulator, which maps each received symbol to the nearest symbol of the \mbox{$M$-QAM} constellation and generates $\hat{\mathbf{s}}\in\mathcal{S}^L$. Given that every symbol in $\hat{\mathbf{s}}$ conveys $\log_2\!M$ bits, the output of the demodulator is a sequence of $n=L\log_2\!M$ bits, denoted by $\mathbf{y}\in\{0,1\}^n$.

The $n$-bit sequence $\mathbf{y}$ is a potentially erroneous copy of codeword $\mathbf{x}$, and can be expressed as $\mathbf{y}=\mathbf{x}\oplus\mathbf{e}$, where $\oplus$ denotes modulo-$2$ addition and $\mathbf{e}\in\{0,1\}^n$, referred to as the \textit{error vector}, represents additive noise. GRAND attempts to determine $\mathbf{x}$ by estimating $\mathbf{e}$~\cite{Duffy2019}. Prospective error vectors, also known as \textit{error patterns}, are ordered in increasing Hamming weight. Each error pattern is subtracted from sequence $\mathbf{y}$ and the result is the estimated codeword if it belongs to \mbox{code-book} $\mathcal{C}$. In other words, if $\hat{\mathbf{e}}$ is the first error pattern in the ordered list of error patterns that satisfies $\mathbf{y}\oplus\hat{\mathbf{e}}\in\mathcal{C}$, then  $\hat{\mathbf{x}}=\mathbf{y}\oplus\hat{\mathbf{e}}$. The $(n-k)\times n$ parity-check matrix $\mathbf{H}$ of the $[n,k]$ block code can be used to verify membership of $\mathbf{y}\oplus\hat{\mathbf{e}}$ in $\mathcal{C}$. Given that \mbox{$\mathbf{G}\mathbf{H}^\mathrm{T}=\mathbf{0}$}, where $\mathbf{H}^\mathrm{T}$ is the transpose of $\mathbf{H}$, the sequence $\mathbf{y}\oplus\hat{\mathbf{e}}$ is a codeword in $\mathcal{C}$ if and only if $\bigl(\mathbf{y}\oplus\hat{\mathbf{e}}\bigr)\mathbf{H}^\mathrm{T}=\mathbf{0}$ \cite{Richardson2008}. To curtail computational complexity without inordinately compromising the error correction capability of GRAND, the Hamming weight of the error patterns in the ordered list could be constrained by an upper limit, termed the \textit{abandonment threshold} $w_\mathrm{th}$~\cite{Duffy2019}. If $w_\mathrm{th}$ is set to no more than half the minimum Hamming distance of the $[n,k]$ block code, the error correction capability of GRAND will be similar to or better than that of existing code-specific decoders~\cite{An2021}.

The concatenation of $M$-QAM demodulation and GRAND results in a hard detection and decoding scheme that operates at the bit level; GRAND, in its current form, neither exploits the structure of the $M$-QAM constellation nor leverages channel state information. Ordering of the error patterns should be guided by the Euclidean and Hamming distances of the hard-detected symbols to neighboring symbols in the constellation, the channel noise and fading. In the following sections, knowledge of the $M$-QAM constellation and the channel state information is integrated in the ordering process of error patterns, and \textit{symbol-level} GRAND is developed and contrasted to the original GRAND~\cite{Duffy2019}, referred to as \textit{bit-level} GRAND in the remainder of the letter.

% ---------- Analysis ----------

\section{Symbol-Level GRAND}
\label{sec:SL_GRAND}

\begin{figure}[t]
\centering
\includegraphics[height=7cm, keepaspectratio]{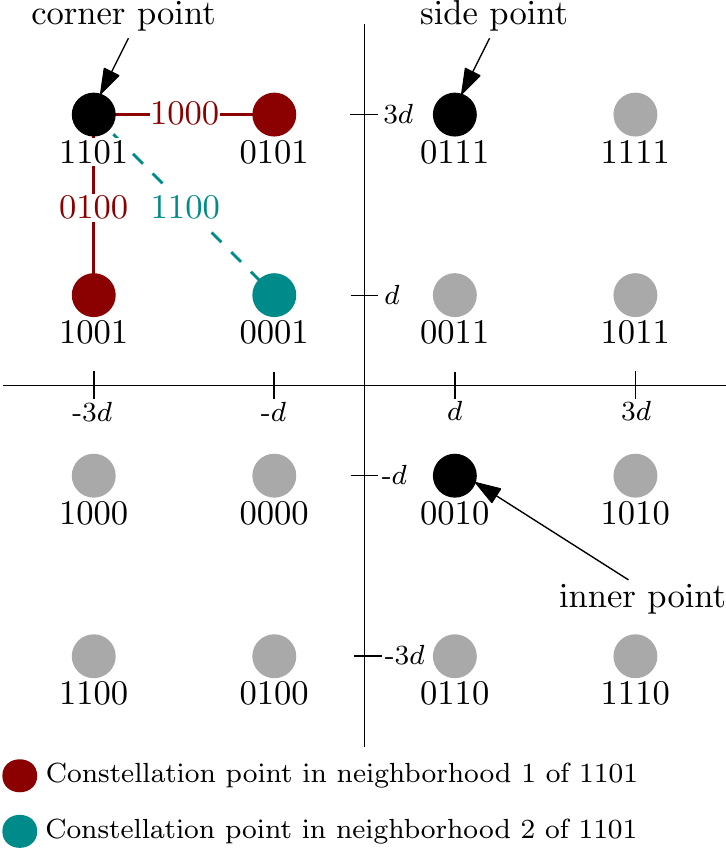}
\caption{Constellation diagram of $16$-QAM. Representatives of corner, side and inner points are shown in black. The nearest neighbors of $1101$ are shown in dark red (neighborhood~1), while the next-nearest neighbor of $1101$ is shown in dark cyan (neighborhood~2). The same color-coding scheme has been used to depict the error strings between $1101$ and each neighboring point.}
\label{fig:16qam_hoods}
%\vspace{-2.6pt}
\end{figure}

The $n$-bit codeword $\mathbf{x}$ defined in Section~\ref{sec:system_model} can be expressed as $\mathbf{x}=(\mathbf{x}_i)_{i=1}^{L}$, where $\mathbf{x}_i$ is the $i$-th string of $\log_2\!M$ bits mapped onto symbol $s_i\in\mathcal{S}$ of $M$-QAM. The correspondence between all possible values of $\mathbf{x}_i$ and the symbols of \mbox{$M$-QAM}, when Gray coding is used, is depicted in Fig.~\ref{fig:16qam_hoods}. The $M$ symbols compose a two-dimensional constellation of equally spaced points, labeled by the mapped values from $\{0,1\}^{\log_2\!M}$. The Euclidean distance between two adjacent points along one dimension is $2d$, where $d$ is given by~\cite{Cho2002, Lu1999}
\begin{equation}
\label{eq:d_def}
d = \sqrt{\frac{3 (\log_2\!M)(k/n)E_b}{2(M-1)}}.
\end{equation}
This distance ensures that the average squared distance of a point from the origin or, equivalently, the average energy per $M$-QAM symbol is $(\log_2\!M)(k/n)E_b$, as stated in Section~\ref{sec:system_model}.

Let $\mathbf{x}_i$ be mapped onto a point of the constellation diagram. We define two neighborhoods around the point with label $\mathbf{x}_i$; \textit{neighborhood 1} encloses points located at Euclidean distance $2d$ from $\mathbf{x}_i$, while
\textit{neighborhood 2} includes points located at Euclidean distance $2\sqrt{2}d$ from $\mathbf{x}_i$. We denote by $\mathcal{E}_1(\mathbf{x}_i)$ and $\mathcal{E}_2(\mathbf{x}_i)$ the sets of all \textit{error strings} obtained by \mbox{modulo-$2$} addition of $\mathbf{x}_i$ with the members of neighborhoods $1$ and $2$, respectively. For example, as shown in Fig.~\ref{fig:16qam_hoods}, neighborhood~$1$ of $\mathbf{x}_i=1101$ consists of the points with labels $0101$ and $1001$; these two points generate the error strings $1101\oplus 0101=1000$ and $1101\oplus 1001=0100$, therefore $\mathcal{E}_1(1101)=\{1000, 0100\}$. Neighborhood~$2$ of $1101$ comprises only one point with label $0001$, hence $\mathcal{E}_2(1101)=\{1100\}$. As a result of Gray coding, the Hamming weight of all members of $\mathcal{E}_1(\mathbf{x}_i)$, for any value of $\mathbf{x}_i$, is $1$. Similarly, the Hamming weight of all members of $\mathcal{E}_2(\mathbf{x}_i)$, for any $\mathbf{x}_i$, is $2$. The cardinalities of $\mathcal{E}_1(\mathbf{x}_i)$ and $\mathcal{E}_2(\mathbf{x}_i)$ depend on the location of the point with label $\mathbf{x}_i$ in the constellation. As illustrated in Fig.~\ref{fig:16qam_hoods}, the square $M$-QAM constellation is composed of \textit{corner} points, \textit{side} points and \textit{inner} points. If $\mathbf{x}_i$ is mapped onto a corner, side or inner point, the number of elements in $\mathcal{E}_1(\mathbf{x}_i)$ is $2$, $3$ or $4$, respectively, while the size of $\mathcal{E}_2(\mathbf{x}_i)$ is $1$, $2$ or $4$, respectively.

At the receiver, the $n$-bit sequence $\mathbf{y}$ at the output of the demodulator can also be written as a sequence of $L$ strings, that is, $\mathbf{y}=(\mathbf{y}_i)_{i=1}^{L}$, where $\mathbf{y}_i$ is a string of $\log_2\!M$ bits that corresponds to a point in the $M$-QAM constellation diagram. \textit{Bit-level} GRAND generates and tests error patterns in order of likelihood until an error pattern $\hat{\mathbf{e}}$ that satisfies $\mathbf{y}\oplus\hat{\mathbf{e}}\in\mathcal{C}$ is found. The likelihood of an error pattern is taken to be inversely proportional to its Hamming weight. In the proposed \textit{symbol-level} GRAND, the requirement for $\mathbf{y}\oplus\hat{\mathbf{e}}\in\mathcal{C}$ remains in place but is expressed as $(\mathbf{y}_i\oplus\hat{\mathbf{e}}_i)_{i=1}^{L}\in\mathcal{C}$, where $\hat{\mathbf{e}}_i$ is the \mbox{$i$-th} error string of the error pattern $\hat{\mathbf{e}}$, and $\mathbf{y}_i\oplus\hat{\mathbf{e}}_i$ is a string that corresponds to a point in the constellation diagram. Given the structure of the \mbox{$M$-QAM} constellation, $\hat{\mathbf{e}}_i$ will be a member of either $\mathcal{E}_1(\mathbf{y}_i)$ or $\mathcal{E}_2(\mathbf{y}_i)$ with high probability, or will be a string of zeros denoted by $\mathbf{0}$. Therefore, in contrast to bit-level GRAND, symbol-level GRAND does not need to generate and test every realization of $\hat{\mathbf{e}}$ for an increasing Hamming weight; instead, it generates only realizations of $\hat{\mathbf{e}}$ that are composed of error strings, which belong to sets $\mathcal{E}_1$ and $\mathcal{E}_2$ of the respective strings in $\hat{\mathbf{y}}$, or are equal to $\mathbf{0}$, that is, $(\mathbf{y}_i\oplus\hat{\mathbf{e}}_i)_{i=1}^{L}\in\mathcal{C}$ for $\hat{\mathbf{e}}_i\in\{\mathbf{0}\}\cup\mathcal{E}_1(\mathbf{y}_i)\cup\mathcal{E}_2(\mathbf{y}_i)$.

Henceforth, for simplicity, we say that an error string $\hat{\mathbf{e}}_i$ is \textit{of type} $\mathcal{E}_j$ if $\hat{\mathbf{e}}_i\in\mathcal{E}_j(\mathbf{y}_i)$ for $j=1,2$. Let $P(L_1, L_2)$ denote the probability of an error pattern $\hat{\mathbf{e}}$ being composed of $L_1$ error strings of type $\mathcal{E}_1$, $L_2$ error strings of type $\mathcal{E}_2$, and $L-L_1-L_2$ error strings that contain zeros, for a given fading coefficient $h$ and noise variance $N_0$. The following proposition derives a tight approximation of $P(L_1,L_2)$.
\begin{proposition}
\label{prop:pattern_composed_of_strings}
An error pattern is a sequence of $L$ error strings, each of length $\log_2\!M$ bits. The probability that $L_1+L_2$ of the error strings are non-zero, when $L_1$ of them are of type $\mathcal{E}_1$ and $L_2$ are of type $\mathcal{E}_2$, can be approximated by:
\begin{IEEEeqnarray}{l}
\label{eq:probability_L1_L2}
P(L_1,L_2)\approx\frac{1}{M^L}%
\nonumber%
\\
\times\!\!\!\!\sum_{L_\mathrm{c}+L_\mathrm{s}+L_\mathrm{i}=L}\!\binom{L}{L_\mathrm{c},L_\mathrm{s},L_\mathrm{i}}4^{L_\mathrm{c}+L_\mathrm{s}}\Bigl(\sqrt{M}-2\Bigr)^{L_\mathrm{s}+2L_\mathrm{i}}
\nonumber%
\\
\times\!\!\!\!\sum_{\substack{L_\mathrm{c,e}+L_\mathrm{s,e}+L_\mathrm{i,e}=\,L_1+L_2 \\ L_\mathrm{c,e}\leq L_\mathrm{c} \\ L_\mathrm{s,e}\leq L_\mathrm{s} \\ L_\mathrm{i,e}\leq L_\mathrm{i}}}\;\,\prod_{\substack{\forall\ell\in\mathcal{L} \\ \mathcal{L}=\{\mathrm{c,s,i}\}}}\!\!\!\binom{L_\ell}{L_{\ell,\mathrm{e}}}p^{L_\ell-L_{\ell,\mathrm{e}}}_{\ell,\mathrm{c}}%
\nonumber%
\\
\times\!\!\!\!\sum_{\substack{L_\mathrm{c,e_1}+L_\mathrm{s,e_1}+L_\mathrm{i,e_1}=\,L_1 \\ L_\mathrm{c,e_1}\leq L_\mathrm{c,e} \\ L_\mathrm{s,e_1}\leq L_\mathrm{s,e} \\ L_\mathrm{i,e_1}\leq L_\mathrm{i,e}}}\;\prod_{\forall\ell\in\mathcal{L}}\!\binom{L_{\ell,\mathrm{e}}}{L_{\ell,\mathrm{e_1}}}p^{L_{\ell,\mathrm{e_1}}}_{\ell,\mathrm{e_1}} p^{L_{\ell,\mathrm{e}}-L_{\ell,\mathrm{e_1}}}_{\ell,\mathrm{e_2}}%
\end{IEEEeqnarray}
where $\mathcal{L}=\{\mathrm{c,s,i}\}$ is a set of indices, which signify the three possible locations of a constellation point, i.e., corner ($\mathrm{c}$), side ($\mathrm{s}$) or inner ($\mathrm{i}$) point, as explained in Fig.~\ref{fig:16qam_hoods}. Expressions for the probabilities $p_{\ell,\mathrm{c}}$, $p_{\ell,\mathrm{e}_1}$ and $p_{\ell,\mathrm{e}_2}$, for $\ell\in\mathcal{L}$, which are functions of $h$ and $N_0$, are provided in Table~\ref{tb:hoold_prob}.
\end{proposition}

\begin{table}[!t]
\renewcommand{\arraystretch}{1.3}
\caption{Expressions for the probability terms in \eqref{eq:probability_L1_L2}. \\ Function $Q(z)\triangleq(1/\sqrt{2\pi})\int_{z}^\infty \exp(-t^2/2)\;dt$ is the tail distribution of the standard normal distribution. Variable $d'$ is defined as $d'\triangleq d\lvert h\rvert / \sqrt{N_0/2}$, where $d$ is given in \eqref{eq:d_def}.}%
\centering%
\begin{tabular}{|c||l|}
\hline $p_{\mathrm{c},\mathrm{c}}$    & $(1-Q(d'))^2$\\
\hline $p_{\mathrm{s},\mathrm{c}}$    & $(1-Q(d'))(1-2Q(d'))$\\
\hline $p_{\mathrm{i},\mathrm{c}}$    & $(1-2Q(d'))^2$\\
\hline%
\hline $p_{\mathrm{c},\mathrm{e}_1}$  & $2(1-Q(d'))Q(d')$\\
\hline $p_{\mathrm{s},\mathrm{e}_1}$  & $2(1-Q(d'))Q(d')+(1-2Q(d'))Q(d')$\\
\hline $p_{\mathrm{i},\mathrm{e}_1}$  & $4(1-2Q(d'))Q(d')$\\
\hline
\hline $p_{\mathrm{c},\mathrm{e}_2}$  & $Q^2(d')$\\
\hline $p_{\mathrm{s},\mathrm{e}_2}$  & $2Q^2(d')$\\
\hline $p_{\mathrm{i},\mathrm{e}_2}$  & $4Q^2(d')$\\
\hline
\end{tabular}
\label{tb:hoold_prob}
\end{table}

\begin{figure}[t]
\centering
\includegraphics[height=6.96cm, keepaspectratio]{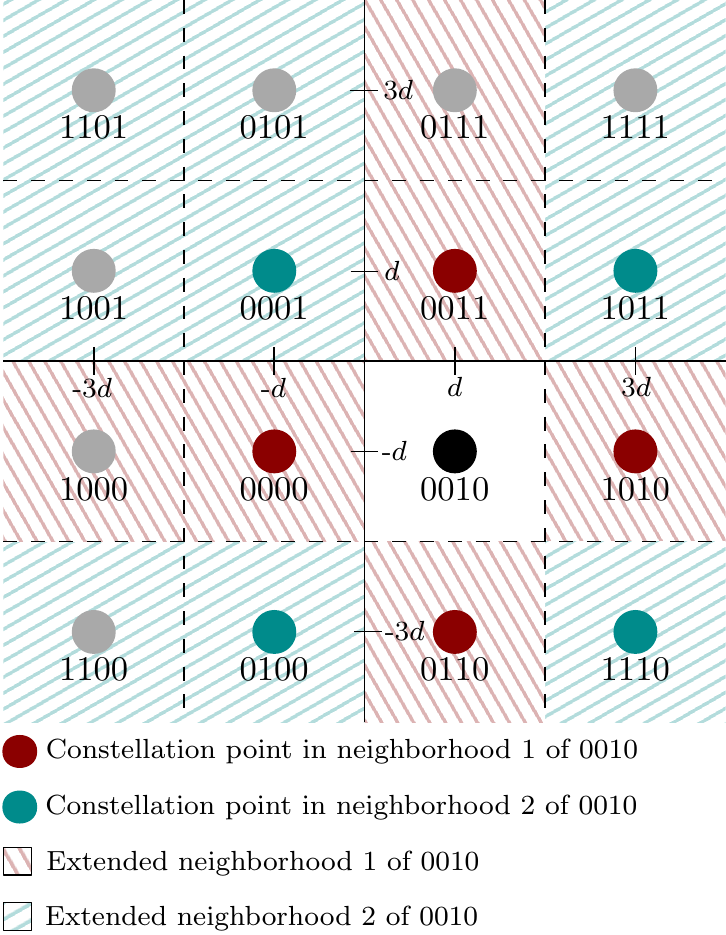}
\caption{Horizontal and vertical lines represent decision boundaries. Points in neighborhoods 1 and 2 of $0010$ are depicted in dark red and dark cyan, respectively, while striped areas indicate the extended neighborhoods 1~and~2.}
\label{fig:16qam_ext_hoods}
\vspace{-4pt}
\end{figure}

\begin{IEEEproof}
To compute $P(L_1,L_2)$, we need to consider the structure of sequences of $L$ error strings. A sequence will contain $L_\mathrm{c}$, $L_\mathrm{s}$ and $L_\mathrm{i}$ error strings that are specific to the neighborhoods of corner, side and inner points, respectively. The $M$-QAM constellation consists of $4$ corner points, $4(\sqrt{M}-2)$ side points and $(\sqrt{M}-2)^2$ inner points. Hence, there exist \mbox{$4^{L_\mathrm{c}}(4(\sqrt{M}-2))^{L_\mathrm{s}}(\sqrt{M}-2)^{2L_\mathrm{i}}$} unique sequences for fixed values of $L_\mathrm{c}$, $L_\mathrm{s}$ and $L_\mathrm{i}$. If we take the sum over all possible values of $L_\mathrm{c}$, $L_\mathrm{s}$ and $L_\mathrm{i}$, provided that $L_\mathrm{c}+L_\mathrm{s}+L_\mathrm{i}=L$, we obtain the second line of \eqref{eq:probability_L1_L2}. Division by $M^L$, which is the number of all possible sequences regardless of their structure, is necessary to ensure that the outcome will be a probability distribution, as shown in the first line of \eqref{eq:probability_L1_L2}. Let us now focus on the $L_\mathrm{i}$ error strings in the sequence, which are associated with inner points ($\ell=\mathrm{i}$). Of them, $L_\mathrm{i,e}$ error strings will correspond to symbols received in error and will be non-zero; $L_\mathrm{i,e_1}$ non-zero error strings will be of type $\mathcal{E}_1$ with probability $p^{L_\mathrm{i,e_1}}_\mathrm{i,e_1}$, while the remaining $L_\mathrm{i,e}-L_\mathrm{i,e_1}$ non-zero error strings will be of type $\mathcal{E}_2$ with probability $p^{L_\mathrm{i,e}-L_\mathrm{i,e_1}}_\mathrm{i,e_2}$. The sequence will also contain $L_\mathrm{i}-L_\mathrm{i,e}$ error strings that are composed of zeros only, as they correspond to symbols that have been received correctly with probability $p^{L_\mathrm{i}-L_\mathrm{i,e}}_\mathrm{i,c}$. Expressions for $p_\mathrm{i,c}$, $p_\mathrm{i,e_1}$ and $p_\mathrm{i,e_2}$ can be found in Table~\ref{tb:hoold_prob}. Derivation of the expressions in Table~\ref{tb:hoold_prob} considered the extended neighborhoods, shown in Fig.~\ref{fig:16qam_ext_hoods} for the case of an inner point, instead of the nearest and next-nearest neighbors. For this reason, expression \eqref{eq:probability_L1_L2} is an approximation that becomes tighter as the channel conditions improve. Although we focused on inner points, the same reasoning can be applied to error strings in the sequence that are associated with corner points ($\ell=\mathrm{c}$) and side points ($\ell=\mathrm{s}$). If we take the sum over all possible values of $L_\mathrm{\ell,e}$ and $L_\mathrm{\ell,e_1}$, for every $\ell\in\mathcal{L}$, provided that the total number of non-zero error strings is $L_1+L_2$ and the total number of non-zero error strings of type $\mathcal{E}_1$ is $L_1$, we obtain the third and fourth lines of \eqref{eq:probability_L1_L2}.
\end{IEEEproof}
\vspace{3pt}

Symbol-level GRAND uses \eqref{eq:probability_L1_L2} to evaluate $P(L_1, L_2)$ for $L_1=0,\ldots,L$ and $L_2=0,\ldots,L-L_1$, where \mbox{$L_1+L_2>0$}, and arranges the obtained probability values in descending order. If $P(L^{*}_1, L^{*}_2)$ is the $i$-th probability value in the ordered list, $L^{*}_1$ and  $L^{*}_2$ are assigned to the entries of the $i$-th row of an $(L(L+3)/2)\times 2$ lookup table $\mathbf{A}=[a_{i,j}]$, that is, $a_{i,1}=L^{*}_1$ and $a_{i,2}=L^{*}_2$. Consequently, the relationship between the entries of consecutive rows of $\mathbf{A}$ is $P(a_{i,1},a_{i,2})\geq P(a_{i+1,1},a_{i+1,2})$. For every row $i$ of $\mathbf{A}$, symbol-level GRAND generates all error patterns that consist of $a_{i,1}$ error strings of type $\mathcal{E}_1$ and $a_{i,2}$ error strings of type $\mathcal{E}_2$. As the algorithm moves from the top row toward the bottom row of  $\mathbf{A}$, the likelihood of the generated error patterns reduces. Symbol-level GRAND terminates when an error pattern $\hat{\mathbf{e}}$ that meets the requirement for $\mathbf{y}\oplus\hat{\mathbf{e}}\in\mathcal{C}$ is identified. The next section verifies the tightness of \eqref{eq:probability_L1_L2}, and compares symbol-level GRAND with bit-level GRAND in terms of performance and complexity.

%A threshold $\zeta_\mathrm{th}$ could be introduced to trigger termination of the algorithm after the top $\zeta_\mathrm{th}$ rows of $\mathbf{A}$ have been considered, even if a suitable error pattern has not been identified.

% ---------- Results ----------

\section{Results and Discussion}
\label{sec:results}

\begin{figure}[t]
\centering
\includegraphics[width=1\columnwidth]{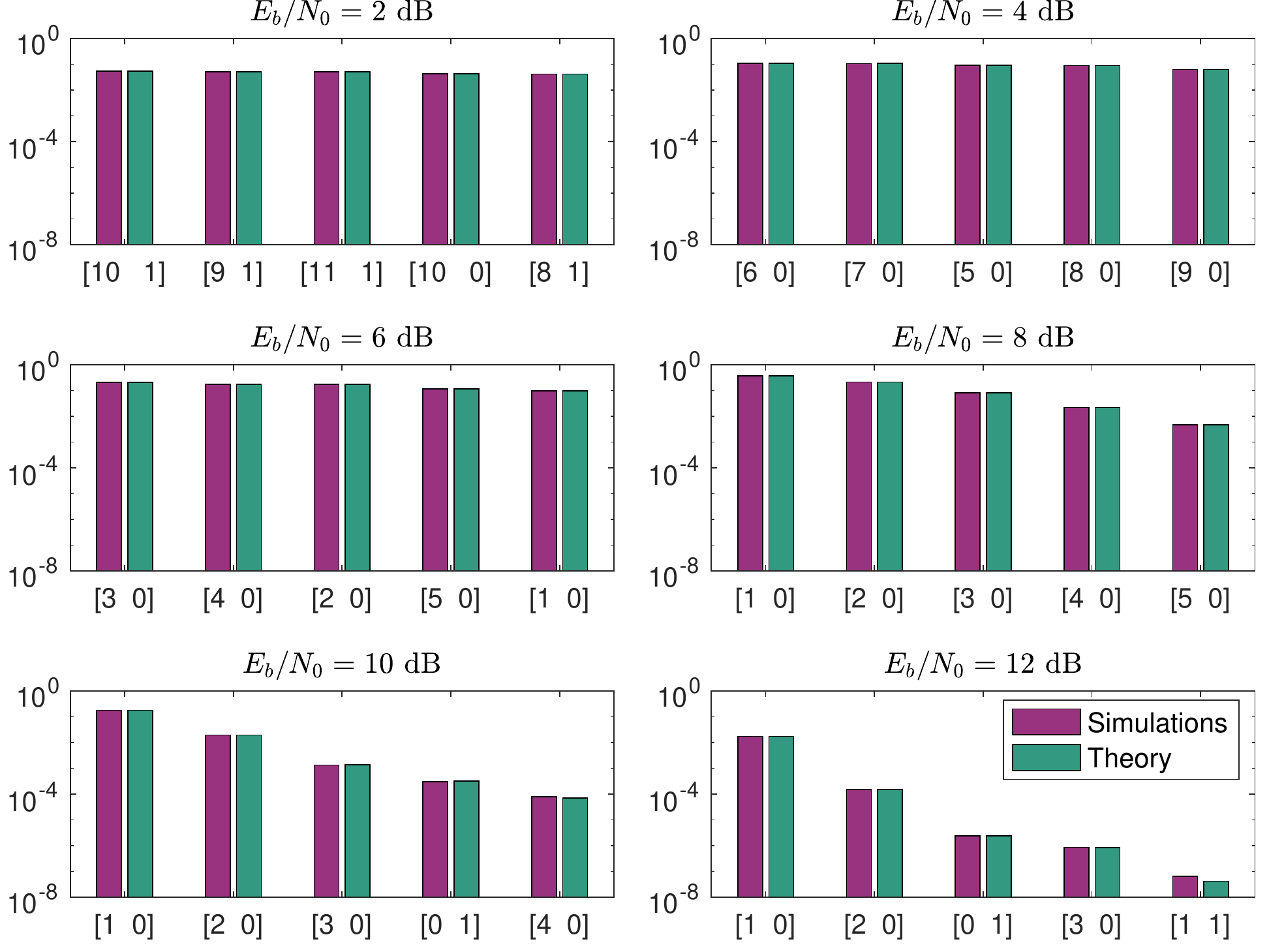}
\caption{Structures of error patterns arranged in order of likelihood for different $E_b/N_0$ values on AWGN. A structure $[L_1\;L_2]$ on the horizontal axis of any subplot represents error patterns containing $L_1$ type-$\mathcal{E}_1$ and $L_2$ type-$\mathcal{E}_2$ error strings, which occur with probability $P(L_1,L_2)$ shown on the vertical axis.}
\label{fig:ordered_error_types}
%\vspace{-2.6pt}
\end{figure}

In order to validate \eqref{eq:probability_L1_L2}, we ran simulations and observed the error patterns at the output of a $16$-QAM demodulator for received sequences impaired by additive noise only, i.e., $h=1$ in \eqref{eq:channel_in_out}. Given that coding does not influence $P(L_1, L_2)$ in \eqref{eq:probability_L1_L2}, we  considered uncoded $16$-QAM, where \mbox{$k=n=128$}. Fig.~\ref{fig:ordered_error_types} compares the frequencies of the five most likely structures of error patterns, which were measured through simulations, with their theoretical occurrence probabilities, as predicted by \eqref{eq:probability_L1_L2}, for various $E_b/N_0$ values. The structure of an error pattern composed of $L_1$ type-$\mathcal{E}_1$ and $L_2$ type-$\mathcal{E}_2$ error strings has been expressed as $[L_1\,\,L_2]$ on the horizontal axis of each subplot in Fig.~\ref{fig:ordered_error_types}. Note that the Hamming weight of an error pattern with structure $[L_1\,\,L_2]$ is  $L_1+2L_2$ for Gray-coded QAM. The vertical axis of each subplot displays the theoretical and measured values of $P(L_1, L_2)$ for each pattern structure. Fig.~\ref{fig:ordered_error_types} demonstrates that theoretical predictions match simulation results, therefore \eqref{eq:probability_L1_L2} is a tight approximation of $P(L_1, L_2)$. As expected, error patterns that contain error strings of type $\mathcal{E}_1$ become dominant for increasing $E_b/N_0$ values. Nevertheless, structures that incorporate error strings of type $\mathcal{E}_2$ continue to appear among the five most likely structures at high $E_b/N_0$ values. Fig.~\ref{fig:ordered_error_types} also confirms that the structure of error patterns is more pivotal to their likelihood than their Hamming weight. For example, when $E_b/N_0\!=\!10$ dB, likelihood drops as we move from one pattern structure to the next, but the Hamming weight does not necessarily increase. %Similar trends can be noticed in other subplots too.

\begin{figure}[t]
\centering
\includegraphics[width=0.99\columnwidth]{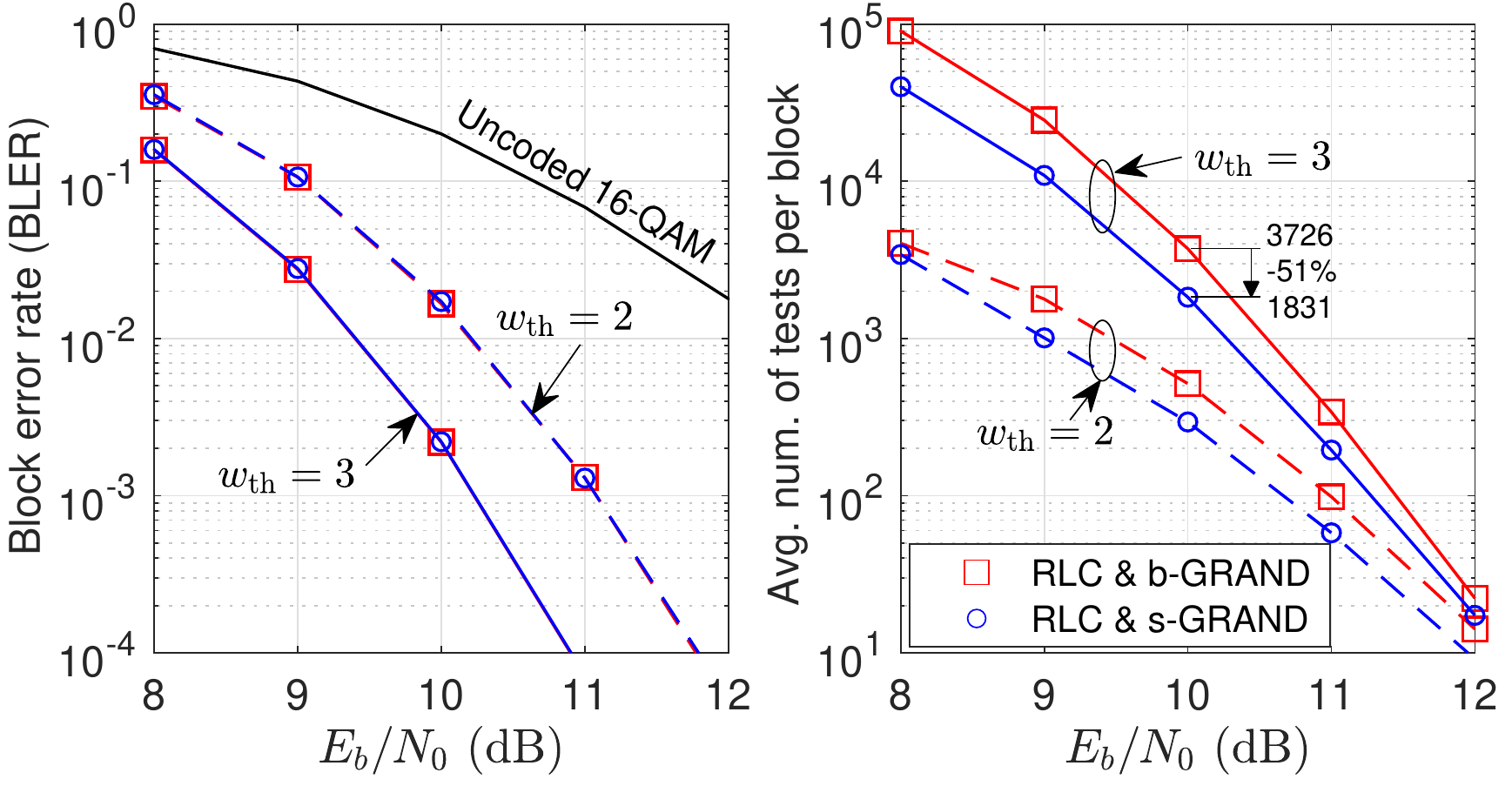}
\caption{BLER and average number of tests per block, as functions of $E_b/N_0$, for bit-level GRAND (b-GRAND) and symbol-level GRAND (s-GRAND), when RLC$[128, 103]$ is used with $16$-QAM. Transmission over \textit{AWGN} is considered. The abandonment threshold is set to $w_\mathrm{th}=2$ and $w_\mathrm{th}=3$.}
\label{fig:16qam_AWGN_sims}
%\vspace{-2.6pt}
\end{figure}

\begin{figure}[t]
\centering
\includegraphics[width=0.99\columnwidth]{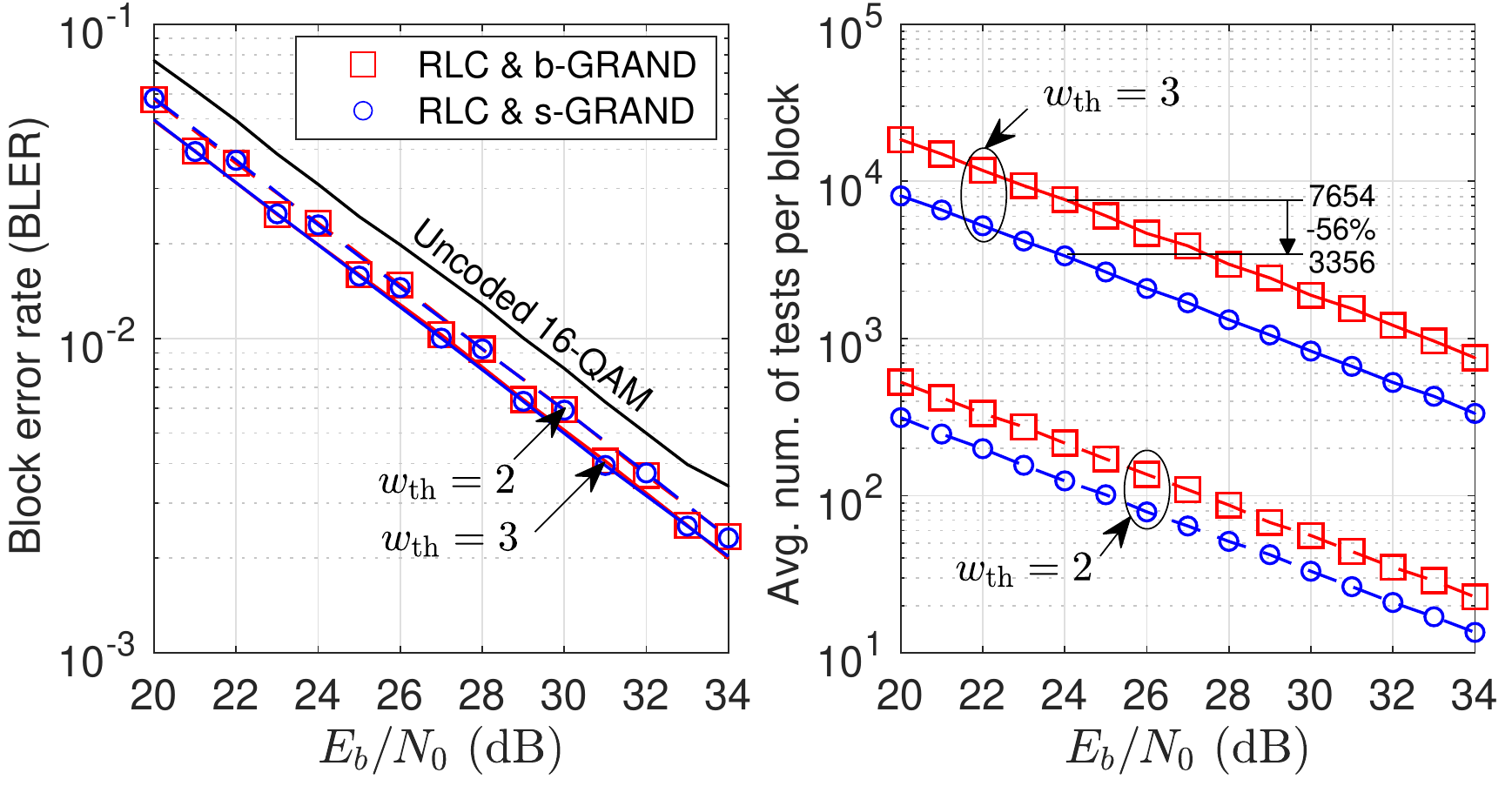}
\caption{BLER and average number of tests per block, as functions of $E_b/N_0$, for bit-level GRAND (b-GRAND) and symbol-level GRAND (s-GRAND), when RLC$[128, 103]$ is used with $16$-QAM on a \textit{Rayleigh fading} channel. The abandonment threshold is set to $w_\mathrm{th}=2$ and $w_\mathrm{th}=3$.}
\label{fig:16qam_Rayleigh_sims}
%\vspace{-2.6pt}
\end{figure}

As \eqref{eq:probability_L1_L2} drives the ordering of error patterns in symbol-level GRAND, a comparison with bit-level GRAND will reveal the impact of the proposed approach on performance and decoding computational complexity. RLC with $k=103$ and $n=128$ has been combined with $16$-QAM at the transmitter. At the receiver, the decoding algorithms consider error patterns of Hamming weight up to $w_\mathrm{th}$, where $w_\mathrm{th}\in\{2,3\}$. Given that the estimated codeword $\hat{\mathbf{x}}$ at the output of the decoder is often referred to as a \textit{block}, the block error rate (BLER) is used as a measure of performance. The number of error patterns that are tested, on average, until the decoder generates an estimate of the transmitted codeword, is adopted as a measure of complexity. Simulation results for AWGN and Rayleigh fading channels are presented in Fig.~\ref{fig:16qam_AWGN_sims} and Fig.~\ref{fig:16qam_Rayleigh_sims}, respectively.

Fig.~\ref{fig:16qam_AWGN_sims} and Fig.~\ref{fig:16qam_Rayleigh_sims} demonstrate that symbol-level GRAND causes no adverse effects on BLER and perfectly matches the BLER of bit-level GRAND. More importantly, both figures establish that symbol-level GRAND converges to a solution faster than bit-level GRAND. In particular, if the channel is impaired by AWGN, symbol-level GRAND tests $15\%$ to $43\%$ fewer error patterns than bit-level GRAND before an estimate of the transmitted codeword is generated for $E_b/N_0\in[8,12]$ and $w_\mathrm{th}=2$. For $w_\mathrm{th}=3$, the drop in the average number of tests is in the range between $22\%$ and $56\%$, as can be seen in Fig.~\ref{fig:16qam_AWGN_sims}. In the case of Rayleigh fading, a switch from bit-level GRAND to symbol-level GRAND reduces the number of tests by $\sim 40\%$ for $w_\mathrm{th}=2$, and by $\sim 56\%$ for $w_\mathrm{th}=3$ when $E_b/N_0\in[20,34]$, as shown in Fig.~\ref{fig:16qam_Rayleigh_sims}.

Symbol-level GRAND offers a notable complexity advantage over bit-level GRAND, albeit with a small increase in memory requirements. A lookup table that contains the most likely structures of error patterns for a range of $E_b/N_0$ values should be constructed, as can be inferred from Fig.~\ref{fig:ordered_error_types}. If structures of the form $[L_1\;L_2]$ that satisfy $0<L_1+2L_2\leq w_\mathrm{th}$ are only considered, then $L_1$ and $L_2$ take values in the ranges $0\leq L_1\leq w_\mathrm{th}$ and $0\leq L_2\leq \lfloor w_\mathrm{th}/2 \rfloor$, where $\lfloor t \rfloor$ denotes the integer part of $t$. Thus, storage of a structure $[L_1\;L_2]$ requires
\begin{equation}
\lambda = \lceil \log_2(w_\mathrm{th}+1) \rceil + \lceil \log_2(\lfloor w_\mathrm{th}/2\rfloor+1) \rceil\;\text{bits}
\label{eq:bits_per_strct}
\end{equation}where $\lceil t \rceil$ = $\lfloor t \rfloor + 1$ if $t>\lfloor t \rfloor$, otherwise $\lceil t \rceil = t$. If only the $\upsilon$ most likely structures for each $E_b/N_0$ value are stored, and $\tau$ values of $E_b/N_0$ are required, the memory size of the lookup table will be $\lambda\upsilon\tau$ bits. For example, the lookup table that was used to obtain the simulation results of symbol-level GRAND for $w_\mathrm{th}=3$ in Fig.~\ref{fig:16qam_Rayleigh_sims} contained $1995$ bits. This is because, according to \eqref{eq:bits_per_strct}, $\lambda=3$ bits per structure of error patterns are needed, the $\upsilon=5$ most likely structures per $E_b/N_0$ value were stored, and $\tau=133$ evenly spaced values of $E_b/N_0$ between $0$~dB and $33$~dB were considered. The size of the lookup table could be further reduced if fewer, but unevenly spaced, $E_b/N_0$ values were recorded to account for the fact that changes in the ordered list of $\upsilon$ error structures occur less frequently as $E_b/N_0$ increases; for high $E_b/N_0$ values (greater than $33$~dB, in our example) no changes in the ordered list are observed.

% ---------- Conclusion ----------

\section{Conclusion}
\label{sec:conclusion}

This letter introduced symbol-level GRAND, a variant of bit-level GRAND that takes into consideration the adopted modulation scheme and the fading coefficients in the search of error patterns that are used in the estimation of transmitted codewords. Simulation results established that symbol-level GRAND requires a marginal increase in memory allocation but offers a notable computational complexity advantage over bit-level GRAND. When RLC$[128,103]$ is combined with \mbox{$16$-QAM} on Rayleigh fading channels, symbol-level GRAND tests up to $56\%$ fewer error patterns than bit-level GRAND and requires less than $2000$ bits (250 bytes) of additional memory.

% ---------- References ----------

\bibliographystyle{IEEEtran}
\bibliography{IEEEabrv, references}

\end{document}